\documentstyle[prl,aps,preprint]{revtex}
\large

\begin{document}
\draft

\title{Strong coupling probe for the Kardar-Parisi-Zhang equation}
\author{T.J. Newman$^{(1)}$\cite{na} and Harald Kallabis$^{(2)}$}
\address{$^{(1)}$Institut f\"ur Theoretische Physik,
Universit\"at zu K\"oln, D-50937 K\"oln, Germany \\
$^{(2)}$H\"ochstleistungsrechenzentrum,
Forschungszentrum J\"ulich, 52425 J\"ulich, Germany \\ }
\date{\today}
\maketitle
\begin{abstract}
We present an exact solution of the {\it deterministic}
Kardar-Parisi-Zhang (KPZ) equation under the influence of a local
driving force $f$. For substrate dimension $d \le 2$ we recover the well-known
result that for arbitrarily small $f>0$, the interface develops
a non-zero velocity $v(f)$. Novel behaviour is found in the
strong-coupling regime for $d > 2$, in which $f$ must exceed a critical force
$f_c$ in order to drive the interface with constant velocity.
We find $v(f) \sim (f-f_c)^{\alpha (d) }$ for $f \searrow f_{c}$.
In particular, the exponent $\alpha (d) = 2/(d-2)$ for $2<d<4$, but saturates
at $\alpha(d)=1$ for $d>4$, indicating
that for this simple problem, there exists a finite upper critical
dimension $d_u=4$. For $d>2$ the surface distortion caused by the
applied force scales logarithmically with distance within a critical
radius $R_{c} \sim (f-f_{c})^{-\nu(d)}$, where
$\nu(d) = \alpha (d)/2$. Connections between these results, and the
critical properties of the weak/strong-coupling transition in the noisy
KPZ equation are pursued.
\end{abstract}
\vspace{5mm}

\newpage

\section{Introduction}

This article is concerned with a problem which
arises in the study of the Kardar-Parisi-Zhang (KPZ) equation\cite{kpz}
of interface growth\cite{rev1}\cite{rev2}. Namely, we consider the
deterministic version of the KPZ equation under the influence of a
local, constant force. Although this problem is in principle rather simple,
a detailed investigation has not been carried out to our knowledge. The
results, although interesting in their own right, assume greater significance
when related to the weak/strong-coupling transition in the noisy KPZ equation.
Before entering into further details of this problem, we shall
give a brief overview of the key properties of the KPZ equation.
In, by now, standard notation this
equation for the evolution of the interface height $h({\bf x},t)$
(above a $d$-dimensional substrate) takes the form
\begin{equation}
\label{e1}
\partial_t h = \nu \nabla ^2 h + (\lambda /2)(\nabla h)^{2} + \eta
\end{equation}
where $\eta ({\bf x},t)$ is a white noise, generally taken to be
gaussian distributed.
The Cole-Hopf transformation $h({\bf x},t) = (2\nu/\lambda )\ln (w({\bf x},t))$
yields a linear equation for $w$ albeit with multiplicative
noise:
\begin{equation}
\label{e2}
\partial_t w = \nu \nabla ^2 w + (\lambda / 2\nu)w\eta  .
\end{equation}
This equation may be re-expressed by means of the Feynman-Kac formula,
such that $w$ represents the restricted partition function of a directed
polymer in the presence of quenched random defects. Naturally, the above
equation may also be interpreted as an imaginary-time Schr\"odinger equation
for a quantum mechanical (QM) particle in a time-dependent
random potential $(-\eta )$.

It is well-known\cite{kpz} that $d=2$ plays the role of a lower critical
dimension.
For $d \le 2$, the interface will be asymptotically (i.e. for large times)
rough for arbitrarily small $\lambda $. For $d>2$ there exists a
transition between asymptotically smooth and rough phases, depending on
the value of $\lambda $. To access the rough (strong-coupling) phase,
$\lambda $ must exceed a critical value $\lambda _c$. Amongst the most
important questions in this field are: i) what are
the scaling exponents associated with the strong-coupling phase; ii) does
there exist an upper critical dimension $d_u$, above which these exponents
take on mean-field values; iii) what are the critical properties
of the weak/strong-coupling transition? There are no
answers to the first question, bar predictions from extensive numerical
work (except for the case $d=1$ where exact analytic results are
available \cite{rev1}).
As regards the second question, the community seems split between
beliefs in $d_u = 4$ and $d_u = \infty$. The former belief is based
primarily on the predictions of various
self-consistent `mode-coupling' theories (see for example \cite{mam}),
whilst the latter belief arises
from numerical evidence for $d$-dependent strong-coupling exponents\cite{rev2}
(for all accessible $d \stackrel{<}{\scriptstyle\sim } 7$).
With regard to the final question, a recent
two-loop renormalization group calculation\cite{ft} (formulated as an
$\epsilon$-expansion, with $\epsilon = d-2$)
predicts that at the transition, the roughness exponent
$\chi = 0 + O(\epsilon ^{3})$ (to be interpreted as logarithmic roughness),
whilst the correlation length exponent has the
form ${\tilde \nu} = (1/\epsilon) + O(\epsilon^{3})$. For the one-loop
versions of these results, see Ref. \cite{nt}.

We now concentrate on a much simpler problem.
As mentioned before, we retain the essential
non-linearity of the KPZ equation, but replace the noise $\eta $ by a
local applied force: $\eta ({\bf x},t) = f_{0} \delta ^{d}({\bf x})$. In
directed polymer language this corresponds to replacing the bulk disorder
by a single constant energy columnar defect. In terms of the QM analogue,
we have replaced the random potential by a delta-function potential with
amplitude $(-f_{0})$.
There exists a well-known result\cite{qm}, which is most
easily stated in QM language: for $d \le 2$ the particle will have a
bound state for arbitrarily small $f_{0}>0$,
whereas for $d>2$ there only exists a bound state for $f_{0} > f_{0,c} >0$.
In terms of the interface, these results imply that for $d\le2$ an arbitrarily
small force is sufficient to induce a non-zero velocity to the surface,
whereas for $d>2$ the driving force must exceed a critical value in order
to achieve this. The identification of a lower critical dimension
(along with the properties of the system for $d\le2$) may be obtained from a
perturbative analysis (in terms of $\lambda$, or equivalently $f_{0}$).

In order to explore the properties of the
`binding/unbinding' transition for $d>2$, a non-perturbative analysis is
required --
such an analysis constitutes the body of
the present work. The main results which arise are the following (which we
shall describe in terms of the interface formulation). Whereas for $d<2$ an
arbitrarily small driving force will induce a velocity $v(f_{0})$ to the
interface, for $d>2$ the applied force must exceed a critical value in order
to achieve this. The dependence of the velocity on the excess force
$\delta f_{0} \equiv f_{0}-f_{0,c}$ is found
to be $v \sim (\delta f_{0})^{\alpha (d)}$ where $\alpha (d) = 2/(d-2)$ for
$2<d<4$, but saturates at $\alpha (d) = 1$ for $d>4$. For $d<2$ the applied
force will create a non-linear distortion of the surface over a critical
radius $R_{c} \sim f_{0}^{-1/(2-d)}$, but the amplitude of the distortion
is of order unity. For $d>2$ the distortion scales logarithmically with
distance and exists within a critical radius
$R_{c} \sim (\delta f_{0})^{-\nu(d)}$, where $\nu(d) = \alpha(d)/2$.
Therefore, for $d>2$ the distortion is a macroscopic rupture
of the surface with a logarithmic profile (by `rupture' we mean a
gross distortion of the interface, leaving the continuous
properties of the surface in tact). We shall
leave the interpretation of these results for the final sections of the paper.

The outline of the paper is as follows. In the next section we shall give a
precise definition of the model to be solved, along with some simple
steps which lead to an accessible strong-coupling (i.e. non-perturbative)
analysis of the model. In section 3, we proceed with this analysis and derive
the velocity-force characteristics. The spatial profiles of the surface
distortion for various dimensions are evaluated in section 4. Connections
between the properties of this simple model, and the weak/strong-coupling
transition in the noisy KPZ model are pursued in section 5. We end the paper
with our conclusions, and a description of possible extensions to the present
work.

\section{Definition and reformulation of the model}

Replacing the stochastic source of the noisy KPZ equation with a local applied
force leads us to consider the equation
\begin{equation}
\label{e3}
\partial_t h = \nu \nabla ^2 h + (\lambda /2)(\nabla h)^{2} +
f_{0}\delta ^{d}({\bf x}) .
\end{equation}
As for the noisy problem, we may apply the transformation
$h({\bf x},t) = (2\nu/\lambda )\ln (w({\bf x},t))$ which yields the linear
equation
\begin{equation}
\label{e4}
\partial_t w = \nu \nabla ^2 w +
(\lambda f_{0}/2\nu) \ w \ \delta^{d}({\bf x}).
\end{equation}
Since we are primarily interested in the interface application of this problem,
we choose the initial condition $w({\bf x},0) = 1$, which corresponds to
an initially flat interface ($h=0$). [The natural initial condition for the
directed polymer application is $w({\bf x},0) = \delta ^{d}({\bf x})$. We shall
not consider this problem here.]

We may formally integrate Eq.(\ref{e4}) using
the Green function of the diffusion equation, $g({\bf x},t)
= (4\pi \nu t)^{-d/2}\exp(-x^{2}/4\nu t)$, leading to
\begin{equation}
\label{e5}
w({\bf x},t) = 1 + (\lambda f_{0}/ 2\nu) \int \limits _{0}^{t}
dt' \ g({\bf x},t-t')w({\bf 0},t') .
\end{equation}
Therefore, knowledge of the function $w({\bf 0},t)$ is sufficient to
determine $w$ throughout all space. For convenience we shall define
$\phi (t) \equiv w({\bf 0},t)$. Setting ${\bf x}={\bf 0}$ in the above
equation leads to a Volterra integral equation for $\phi \ $ :
\begin{equation}
\label{e6}
\phi (t) = 1 + f\int \limits _{0}^{t} dt' \ \sigma (t-t') \phi (t')
\end{equation}
where $\sigma (t) = t^{-d/2}$ and we have defined an effective force
$f \equiv (\lambda f_{0}/ 2\nu)(4 \pi \nu)^{-d/2}$. [Such a Volterra equation
may also be derived for the more complicated situation
of a time-dependent force $f(t)$. The case where
$f(t)$ is a stochastic variable is of particular interest in the directed
polymer picture \cite{kl}.]

Clearly this equation is ill-defined for $d>2$, since the kernel is
non-integrable over the range $(0,t)$.
To regularize the equation we have many possibilities.
The most convenient is to smear the delta-function force over a small region
of size $l^{d}$ with a gaussian envelope, i.e. we replace
$f_{0} \delta ^{d}({\bf x})$ by $f_{0} (\pi l^{2})^{-d/2} \exp (-x^{2}/l^{2})$.
The length scale $l$ constitutes the smallest physical scale in the problem.
Assuming that $w({\bf x},t)$ varies slowly over the core region $|{\bf x}|< l$
allows us to rederive the above Volterra equation, where now
$\sigma (t) = (t + l^{2})^{-d/2}$ (after rescaling $l \rightarrow l
= l/(4\nu )$).

Laplace transforming Eq.(\ref{e6}) and using the convolution theorem
yields an explicit solution for this Laplace transform ${\cal L}[\phi (t)]$.
Inverting the transform
gives the result
\begin{equation}
\label{e7}
\phi(t) = {1 \over 2\pi i} \int _{\gamma }  \ {ds \over s} \
{e^{st/l^{2}} \over [1 - l^{2-d} f D(s)] }
\end{equation}
where $\gamma $ is the usual integration
contour along the line $s = c+iy$ where $c$ is a real constant
chosen to lie to the right of any singularities in the complex plane.
The function $D(s)$ is given by
\begin{equation}
\label{e8}
D(s) = \int \limits _{0}^{\infty} dy \ {e^{-sy} \over (1+y)^{d/2}}
\end{equation}
(which may be expressed in terms of the incomplete Gamma function\cite{as}).

We are now in a position to obtain all physical properties of the model
from the analysis of Eq.(\ref{e7}). We stress once again that this
analysis is non-perturbative in $f_{0}$ (or equivalently $\lambda $). All
the results for $d \le 2$ may be obtained by perturbation theory (although
one must evaluate all orders), whereas none of the results concerning
the binding/unbinding transition may be obtained by such a technique for $d>2$.

In the next section we shall concentrate on evaluating the asymptotic
(large time) form of $\phi(t)$ which will enable us to determine whether
the interface has a non-zero velocity. The spatial
profile of the interface distortion requires the evaluation of the convolution
of the diffusion equation Green function with the function $\phi (t)$ (as
illustrated in Eq. (\ref{e5})) - this analysis is described in section 4.

\section{Velocity-force characteristics}

The singularities in the integrand of Eq.(\ref{e7}) consist of a branch cut
along the negative real axis, and (depending on the value of $f$) a pole
in the right-half plane (RHP). The large-time behaviour of $\phi(t)$ will
be dominated by the contribution from the pole (if it exists). One may easily
convince oneself that the existence of this pole is required for the
interface at the forcing site to have a non-zero velocity, i.e. for
$H(t) \sim vt$
(where $H(t) \equiv (\lambda/2\nu )h({\bf 0},t) = \ln \ \phi (t)$.)

By studying the
small-$s$ behaviour of $D(s)$ one may show the following properties of
the integrand in Eq.(\ref{e7}). For $d < 2$ there always exists a pole
in the RHP for $f>0$, which for $f \searrow 0$ is situated at
$s_{p}=f^{2/(2-d)}$. For $d>2$ the pole will only exist in the RHP for
$f >  f_{c} = l^{d-2}(d-2)/2$. For $f \searrow f_{c}$, the
position of the pole is given by $s_{p}=A(d)(f-f_{c})^{2/(d-2)}$ for
$2<d<4$ (where A(d) is a complicated function of $d$), whereas for $d>4$
one has $s_{p} = [(d-4)/(d-2)](f-f_{c})$. This abrupt change of
behaviour for $s_{p}$ at $d=4$ will have interesting consequences.

Evaluating the contribution to $\phi (t)$ from the pole leads us to the
following results for the velocity-force characteristics: i) for $d<2$ one
has for $f \searrow 0$ the relation $v(f) \sim f^{2/(2-d)}$; ii) for
$d>2$, the velocity is zero unless $f$ exceeds $f_{c}$ --
one then has for $f \searrow f_{c}$ the relation
$v(f) \sim (\delta f)^{\alpha (d)}$ where $\delta f \equiv f-f_{c}$ and
\begin{equation}
\label{e9}
\nonumber
\alpha (d) = \left \{
\begin{array}{ll}
\frac{2}{(d-2)}
\hspace{20mm} & 2 < d < 4 \\
1 & d > 4 \ ,
\end{array}
\right.
\end{equation}
which signals the existence of an upper critical dimension $d_{u}=4$.

In the absence of a pole in the RHP, the asymptotics of $\phi (t)$ arise
from the contribution from the branch cut. The results for this case (along
with the $v(f)$ characteristics at the critical dimensions 2 and 4) are
displayed for completeness in Table I.

\section{Spatial profiles}

In this section we wish to evaluate the spatial profile of the
interface for $f > f_{c}$. We define the positive quantity
$\Delta H({\bf x},t) \equiv (\lambda /2\nu)(h({\bf 0},t)-h({\bf x},t))$. Making
use
of the Cole-Hopf transformation, along with Eq. (\ref{e5}), we have (after
rescaling
${\bf x} \rightarrow {\bf x} = {\bf x}/(4\nu)$)
\begin{equation}
\label{e10}
\Delta H = -\ln \left \lbrace {1\over \phi(t)} + {f\over \phi(t)}\int \limits
_{0}^{t}
dt' \ g({\bf x},t'+l^{2}) \ \phi(t-t') \right \rbrace .
\end{equation}
For $f>f_{c}$ and $vt \gg 1$, we have from the previous analysis
$\phi (t) \sim e^{vt}$. In this case, the above expression reduces to
\begin{equation}
\label{e11}
\Delta H = -\ln [fF({\bf x})]
\end{equation}
where
\begin{equation}
\label{e12}
F({\bf x}) =  \int \limits _{0}^{\infty} dt' \ e^{-vt'} \ g({\bf x},t'+l^{2}) .
\end{equation}
This expression is valid for $|{\bf x}| \ll v^{1/2}t$. For this entire range
we see that the height deviation $\Delta h$ is independent of time. For
extremely large distances from the forcing site ($|{\bf x}| \gg v^{1/2}t$),
the influence of the force is insufficient to induce a velocity,
and the surface distortion is essentially zero.

We now concentrate on examining the form of the steady-state profile
given by Eqs.(\ref {e11}) and (\ref {e12}). In fact there are {\it two}
spatial regimes for this profile, separated by the length scale
$R_{c} = v^{-1/2}$ which will emerge as a {\it critical radius}.
We shall denote the regime $R_{c} \ll |{\bf x}| \ll v^{1/2}t$
as region A, and the regime $l \ll |{\bf x}| \ll R_{c}$ as region B.
It is important to stress that for $vt \gg 1$ and $v \ll 1$ (which is
the case when $f \searrow f_{c}$), regions A and B are both large and
well-separated. Region A is less interesting and simpler to analyse, so
we shall focus on this first.

\subsection{$\Delta H$ in region A}

In this region, $F({\bf x})$ may be evaluated using a steepest-descents
treatment. A straight-forward calculation yields the result
\begin{equation}
\label{e13}
\nonumber
\Delta H({\bf x}) = \left \{
\begin{array}{ll}
2|{\bf x}|/R_{c} + {\rm O}\left ( \ln (|{\bf x}|/R_{c}) \right ),
\hspace{25mm} & d<2 \\
2|{\bf x}|/R_{c} +  {\rm O}\left ( \ln (|{\bf x}|/R_{c}) \right )
+ {\rm O}\left ( \ln (R_{c}/l) \right ), & d > 2 \
\end{array}
\right.
\end{equation}
The form of the corrections is of interest. We see that for $d<2$, the
leading term dominates for $|{\bf x}| \gg R_{c}$ as expected. This indicates
that the overall height deviation within the critical radius is of
order unity. For $d>2$ the microscopic length scale enters into the
corrections and the condition for the dominance of the leading term
is now $|{\bf x}| \gg R_{c}\ln(R_{c}/l)$. We are therefore given the
hint that for $d>2$ the overall deviation of the height within the critical
radius may be large (of order $\ln(R_{c}/l)$). To see this explicitly
we now turn our attention to the form of $\Delta H$ for
$|{\bf x}| \ll R_{c}$.

\subsection{$\Delta H$ in region B}

Although the integral determining $F({\bf x})$ appears quite simple,
the analysis in region B is rather complicated. We therefore refer
the reader to Appendix A where this analysis is described in some
detail. The end result for $\Delta H$ in region B takes the form
\begin{equation}
\label{e14}
\nonumber
\Delta H = \left \{
\begin{array}{ll}
c(d)f \ |{\bf x}|^{2-d} + {\rm O}\left ( (|{\bf x}|/R_{c})^{2} \right ),
\hspace{25mm} & d < 2 \\
2f\ln(|{\bf x}|/l) + {\rm O}\left ( f \right ),
\hspace{25mm} & d=2 \\
(d-2)\ln(|{\bf x}|/l) +  {\rm O}\left ( 1 \right ), & d > 2 \
\end{array}
\right.
\end{equation}
where $c(d) = 2\Gamma (d/2)/(2-d)$. [It is instructive for $d=2$ to rewrite
the prefactor of the height deviation exclusively in terms of $R_{c}$ --
making use of the relation $2f = 1/\ln  (R_{c}/l)$].

There are two main features of these results we wish to stress.
First, there exists for all dimensions a critical radius $R_{c}$
within which the surface profile undergoes a distortion from the
linear form shown in Eq.(\ref{e13}). The critical radius scales
with $\delta f \equiv f-f_{c}$ as $R_{c} \sim (\delta f)^{-\nu}$
where $\nu (d) = \alpha (d)/2$. In particular we note that
for $2<d<4$, $\nu(d) = 1/(d-2)$; and for $d>4$, $\nu(d) = 1/2$.

The second important point is that by calculating the overall
height distortion within the critical radius, i.e. $\Delta H(R_{c})$,
we see that
\begin{equation}
\label{e15}
\nonumber
\Delta H(R_{c}) \sim \left \{
\begin{array}{ll}
{\rm O}(1)
\hspace{20mm} & 0 \le d \le 2 \\
\ln (R_{c}/l) & d > 2 \
\end{array}
\right.
\end{equation}
These results confirm the ideas propounded earlier, i.e. for $d \le 2$
the distortion of the surface within the critical radius is negligible,
whereas for $d>2$ the distortion is a macroscopic rupture which
scales logarithmically with distance.

\section{Applications to KPZ weak/strong-coupling transition}

As we mentioned in the Introduction, there are two main
characteristics of the weak/strong coupling transition for the noisy
KPZ equation (which arise from a RG study in powers of $\epsilon =
d-2$) for $d>2$. These are the vanishing of the roughness exponent,
which implies a logarithmic roughness; and the existence of a
correlation length $\xi \sim (g-g_{c})^{-{\tilde \nu}}$, where $g$ is the
dimensionless coupling constant, and ${\tilde \nu} = 1/\epsilon +
O(\epsilon )^{3}$. There is a remarkable correspondence between these
results, and those found in the previous section for the simple model
of a deterministic KPZ interface under a local applied force. To
reiterate these results, we found that for $d>2$, the applied force
induces a logarithmic distortion of the surface within a critical
radius which scales as $R_{c} \sim (f-f_{c})^{-\nu}$ with $\nu =
1/\epsilon $ for $2<d<4$, and $\nu=1/2$ for $d>4$.

The natural question arises: are the critical properties of the noisy
KPZ equation directly related to the simple problem discussed above?
If so, then we have a much easier way to understand the weak/strong
coupling phase transition, and also we may identify an upper critical
dimension of 4 (for the unstable fixed point characterising the
transition).  A quantitative connection between the two sets of
results is beyond our reach. However we shall make several points
which certainly make the connection more plausible. Since we are
interested in the weak/strong coupling transition, we henceforth
restrict our attention to $d>2$.

The coupling constant $g$ in the noisy KPZ equation is proportional to
the noise amplitude $D$ which we shall take to be our control
parameter.  For $D$ well below the critical value, the non-linearity
in the KPZ equation is irrelevant. Thus, the surface behaves as an
Edwards-Wilkinson (EW) interface \cite{ew}, which is smooth. As we
increase the value of D, it becomes more and more likely for small
regions in the interface to experience strong fluctuations due to rare
events in the noise spectrum. If the smooth/rough transition is
triggered by an inherent instability of the interface with respect to strong
fluctuations, it is plausible to assume that such an instability is
localized due to the local nature of rare events in the noise.
Having established this, we may connect the two models by arguing that
these strong local fluctuations resemble an effective force over long
time scales. There are three essential steps in this argument which
we discuss in turn.

First, we must argue that the fluctuations appear as a constant force
over some temporal interval $\tau $. Clearly any fluctuations in the `downward'
direction (i.e. anti-parallel to the direction of positive $h$) will
have no effect since they are completely suppressed by the
non-linearity. We therefore consider a noise path (or noise `history') which
tends on
average to be positive over the interval $\tau $ within some region
$\Lambda $ of size $l^{d}$.  This path may be considered as a force
of magnitude $f$ so
long as it lies between the bounds $f \pm \sigma $. The fluctuations
within these bounds are irrelevant. This fact may be established by
considering the delta-function model Eq.(\ref{e3}) but allowing for
the force to be a stochastic variable. By power counting one may show
that the fluctuations of the force are irrelevant (about the constant
force fixed point) for $d>1$.  For the noise path to exceed $f+\sigma$
is extremely unlikely as the path is already a rare event even to have
entered between these bounds.  The path is much more likely to drop
below $f-\sigma$. This is deemed as switching off the `force'. The
force will be switched on again when the noise path returns within the
bounds. Thus the interval $\tau$ over which the force acts is a random
variable.

Second, we must argue that this random switching off and on of the
effective force is unimportant. We assume that the force has been able
to create some disturbance in the interface over the region $\Lambda$,
before it is switched off. How will the surface evolve? To answer this
question one may simply solve the deterministic KPZ equation in the
absence of any source, but with an initial condition corresponding to
a localized disturbance. The result is that for any disturbance
exceeding a critical height $h_{c} \sim \nu /\lambda$, the centre of
the disturbance decays logarithmically slowly -- it is essentially
frozen. This effect is due to the strong influence of the
non-linearity driving the disturbance upwards against the smoothing
effect of the diffusion term. So during the periods of zero force, the
disturbance in the region $\Lambda $ is frozen and on switching on the
force, the evolution continues as if there were no interruption.

The third and final point is to argue that the effective force remains
localized in the region $\Lambda $. This is in general not the case.
The noise has short-range correlations in space and time and rare
events are equally likely to occur anywhere in the surface. A simple
resolution would be that once the rupture has been formed in $\Lambda $,
then all other rare events in the vicinity may be rotationally averaged
to give an effective force within $\Lambda $. A more subtle point
is the following. Once the logarithmic distortion
centered in $\Lambda$ has been established, it is essentially frozen
during the periods of zero force. Therefore any rare event occurring
in $\Lambda $ will continue the evolution of the distortion. On the
other hand, if the critical height for freezing is not small, then
most rare events will fail to seed such a distortion of the interface,
since the disturbances due to rare events which are of a height less
than $h_{c}$ will simply diffuse away. We therefore have a picture of
extremely strong fluctuations being required to seed a distortion of
height $> h_{c}$, with less strong fluctuations being required to
sustain the evolution of the logarithmic rupture once it has been seeded.

If this rough physical picture is correct then the
correspondence between the local
force model considered here, and the noisy KPZ equation at the
weak/strong-coupling transition is established. The surface morphology
resulting from such a picture is that of a dilute system of
macroscopic distortions, each with a logarithmic profile. This is in
contradistinction to the more conventional view of weak uniform
fluctuations giving rise to logarithmic roughness. Such a difference
in morphology should be discernible from numerical simulations of
models believed to lie in the KPZ universality class.

[A physically plausible connection between these two models may
be drawn also in the directed polymer
representation. The strong-coupling phase of the KPZ model
corresponds to the low-temperature phase of the directed polymer,
in which the polymer is stongly localized onto low-energy
paths, which have characteristic transverse fluctuations. Conversely,
the weak-coupling phase of KPZ corresponds to the high-temperature
phase of the directed polymer in which the polymer is liberated from
the low energy path and performs thermal wandering.
Hence the transition between these two phases for the directed polymer
may be viewed as the localization transition to the lowest energy path
(at this temperature). In this case the polymer is essentially subjected
to a `columnar' defect, although the defect is not of constant
energy, and also has transverse fluctuations. Using similar arguments
to those presented above in the interface language, one can
motivate the idea that i) the energy fluctuations on the column are
irrelevant within some bounds, and ii) the transverse fluctuations
of the column are thermally averaged such that the effective column
is straight on some transverse scale of order $l^{d}$. We believe a more
quantitative connection may be tractable in this directed polymer picture.]

\section{Conclusions}

We have presented a non-perturbative analysis of the deterministic KPZ
equation under the influence of a local driving force f, which for
$d>2$ must exceed a critical value $f_{c}$ in order to induce a
non-zero velocity in the surface. The central result is that for $d>2$
and for $f \searrow f_{c}$, the force creates a macroscopic distortion
of the surface. This distortion has a logarithmic profile and exists
within a critical radius of size $R_{c} \sim (f-f_{c})^{-\nu}$ where
$\nu (d) = 1/(d-2)$ for $2<d<4$ and $\nu(d) = 1/2$ for $d>4$.  In the
penultimate section we offered some physical arguments as to why one
might consider this simple model to underlie the weak/strong-coupling
transition in the noisy KPZ model. The main characteristics of such a
scenario are that i) the logarithmic roughness at the transition is
due to a dilute system of logarithmic distortions, as opposed to a
uniform sea of fluctuations with logarithmic variance, and ii) the
critical radius $R_{c}$ is a physical realization of the correlation
length found from the RG treatment. A more quantitative connection
between these two models is presently beyond our reach, although one future
possibility is to estimate the $d$-dependence of the weak/strong-coupling
phase boundary based on the exact results obtained in this paper for
the local force model, in conjunction with the statistical arguments
presented above.

It would be useful to apply the non-perturbative methods presented
here to the case of a random local force. Such an analysis would
allow one to tighten some of the arguments given in section 5.
This problem is also of substantial interest in its own right as it may be
mapped to a system of two directed polymers with random
contact interactions\cite{kl}\cite{lip}\cite{more} which has a wide
range of applications.

\vspace{10mm}

The authors wish to thank M.A.Moore and L-H. Tang for illuminating
discussions. TJN acknowledges financial support under SFB 341.

\newpage

\appendix
\section{}

In this appendix we analyse the function $F({\bf x})$ in region B,
i.e.  for $l \ll |{\bf x}| \ll R_{c} = v^{-1/2}$, in order to derive
the relations shown in Eq.(\ref{e14}). The function $F({\bf x})$ has
the explicit form (cf.  Eq.(\ref{e12}))
\begin{equation}
\label{a1}
F({\bf x}) = \int \limits _{0}^{\infty} dt \ (t+l^{2})^{-d/2} \ \exp
(-vt) \ \exp [-x^{2}/(t+l^{2})] ,
\end{equation}
which may be re-expressed as
\begin{equation}
\label{a2}
F({\bf x}) = x^{2-d} \ e^{vl^{2}} \ [I_{1}(x)+I_{2}(x)]
\end{equation}
where
\begin{equation}
\label{a3}
I_{1}(x) = \int \limits _{(l/x)^{2}}^{1} dt \ t^{-d/2} \ e^{-at} \
e^{-1/t} ,
\end{equation}
and
\begin{equation}
\label{a4}
I_{2}(x) = \int \limits _{1}^{\infty} dt \ t^{-d/2} \ e^{-at} \
e^{-1/t}
\end{equation}
where we have defined $a \equiv vx^{2} = (x/R_{c})^{2} \ll 1$ for
notational convenience.

The first integral may be simply evaluated to leading order in $a$.
We have
\begin{equation}
\label{a5}
I_{1}(x) = \int \limits _{1}^{\infty} du \ u^{d/2-2} \ e^{-u} + {\rm
  O}(a) + {\rm O}(e^{-(x/l)^{2}}) .
\end{equation}

The leading terms of the second integral may be extracted as follows.
Referring to Eq.(\ref{a4}), we expand the last factor of the integrand
as a power series to give
\begin{equation}
\label{a6}
I_{2}(x) = \sum \limits _{n=0}^{\infty} {(-1)^{n}\over n!}J_{n}(a)
\end{equation}
where
\begin{equation}
\label{a7}
J_{n}(a) = a^{n-1+d/2} \ \int \limits _{a}^{\infty} dt \ t^{-n-d/2} \
e^{-t} .
\end{equation}
Integrating by parts yields the recursion relation
\begin{equation}
\label{a8}
J_{n}(a) = {e^{-a} \over (n-1+d/2)} - {a \over (n-1+d/2)}J_{n-1}(a) .
\end{equation}
For $d<2$, $J_{0}(a)$ has the expansion
\begin{equation}
\label{a9}
J_{0}(a) = \Gamma (1-d/2) \ a^{d/2-1} - (1-d/2)^{-1} + {\rm O}(a)
\end{equation}
where $\Gamma (z)$ is the Gamma function\cite{as}.  This form of
$J_{0}$ implies $J_{1}(a) = 2/d + {\rm O}(a^{d/2})$ and for $n>1$,
$J_{n} = (n-1+d/2)^{-1} + {\rm O}(a)$. Returning to Eq.(\ref{a6}) we
have
\begin{equation}
\label{a10}
I_{2}(x) = \Gamma (1-d/2) \ a^{d/2-1} + C + {\rm O}(a^{d/2})
\end{equation}
where
\begin{equation}
\label{a11}
C = \sum \limits _{n=0}^{\infty} {(-1)^{n} \over n!} (n-1+d/2)^{-1} =
(d/2-1)^{-1} + \int \limits _{0}^{1} du \ u^{d/2-2} \ [e^{-u}-1] .
\end{equation}
Combining Eqs.(\ref{a5}) and (\ref{a10}) then gives for $d<2$
\begin{equation}
\label{a12}
I_{1} + I_{2} = \Gamma (1-d/2) \ a^{d/2-1} - (1-d/2)^{-1}\Gamma (d/2)
+ {\rm O}(a^{d/2}).
\end{equation}
 From Table I we have for $d<2 \ $, $v = [f \ \Gamma
(1-d/2)]^{2/(2-d)}$, which in addition to Eqs.(\ref{a2}) and
(\ref{a12}) gives
\begin{equation}
\label{a13}
fF({\bf x}) = 1 - f(1-d/2)^{-1}\Gamma(d/2) \ |{\bf x}|^{2-d} + {\rm
  O}\left ( (|{\bf x}|/R_{c})^{2} \right ) .
\end{equation}
Combining the above equation with Eq.(\ref{e11}) then yields the first
relation in Eq.(\ref{e14}).

A similar analysis suffices for higher dimensions, which we shall
briefly sketch. For $d=2$ one finds
\begin{equation}
\label{a14}
I_{1} + I_{2} = -\ln (a) - 2\gamma + {\rm O}(a \ln (a)),
\end{equation}
where $\gamma = 0.57721...$ is Euler's constant.  From Table I we have
for $d=2 \ $, $v = l^{-2} \ e^{-1/f}$, which in addition to
Eqs.(\ref{a2}) and (\ref{a14}) gives
\begin{equation}
\label{a15}
fF({\bf x}) = 1 - 2f \ln (|{\bf x}|/l) + O(f) .
\end{equation}
Combining the above equation with Eq.(\ref{e11}) then yields the
second relation in Eq.(\ref{e14}).

Finally for $d>2$ one finds
\begin{equation}
\label{a16}
I_{1} + I_{2} = \Gamma (d/2-1) + {\rm O}(a^{\beta})
\end{equation}
where $\beta = d/2-1$ for $2<d<4$, and $\beta = 1$ for $d>4$.  From
Table I we have for $d>2$, $f \sim f_{c} = (d/2-1)l^{d-2}$, which in
addition to Eqs.(\ref{a2}) and (\ref{a16}) gives
\begin{equation}
\label{a17}
fF({\bf x}) = \Gamma (d/2) \ (l/|{\bf x}|)^{d-2} + {\rm O} \left (
  (|{\bf x}|/R_{c})^{2\beta} (l/|{\bf x}|)^{d-2} \right ) .
\end{equation}
Combining the above equation with Eq.(\ref{e11}) then yields the third
relation in Eq.(\ref{e14}).

\newpage

\noindent
\begin{center}
  {\bf Table I}
\end{center}

\vspace{10mm}

\begin{tabular}{|c||c|c|c|c|c|}\hline
  \raisebox{-1.5ex}[1.5ex] {$d$} & \raisebox{-1.5ex}[1.5ex] {$f_{c}$}
  & \multicolumn{3}{|c|}{$H(t)$} & \raisebox{-1.5ex}[1.5ex] {$v(f)$}
  \\ \cline{3-5} & & $f<f_{c}$ & $f=f_{c}$ & \ $f>f_{c}$ \ & \\ \hline
  \hline

  $0 \le d < 2$ & 0 & \ $-{(2-d)\over 2} \ln \left ( f^{2/(2-d)}t
  \right )$ \ & 0 & $v \, t$ & $c_{1}(d) \ f^{2/(2-d)}$ \\ \hline

  2 & 0 & $-\ln \left ( \ln (t/l^{2}) \right )$ & 0 & $v \, t$ & ${1
    \over l^{2}} \exp (-1/f)$ \\ \hline

  \ $2<d<4$ \ & \ ${(d-2)\over 2}l ^{d-2}$ \ & $-\ln \left ( 1-{f\over
      f_{c}}\right )$ & ${(d-2)\over 2}\ln (t/l^{2})$ & $v \, t$ & \ \
  ${c_{2}(d)\over l^{2}} (\delta \! f/f_{c})^{2/(d-2)}$ \ \ \\ \hline

  4 & $l^{2}$ & $-\ln \left ( 1-{f\over f_{c}}\right )$ & $\ln \left [
    {t/l^{2} \over \ln (t/l^{2} ) } \right ]$ & $v \, t$ & ${1\over
    l^{2}} {(\delta \! f/f_{c}) \over \ln (f_{c}/\delta \! f)}$ \\
  \hline

  $d>4$ & ${(d-2)\over 2}l ^{d-2}$ & $-\ln \left ( 1-{f\over f_{c}}
  \right )$ & $\ln (t/l^{2})$ & $v \, t$ & ${1\over l ^{2}} {(d-4)
    \over 2} \ (\delta \! f/f_{c}) $ \\ \hline
\end{tabular}

\vspace{70mm}
\noindent
{\bf Caption for Table I}

\vspace{10mm}

\noindent
Asymptotic behaviour of $H(t)$ -- the interface height at the forcing
site -- as a function of dimension $d$ and applied force $f$.  The
symbol $\delta \! f$ is the excess force: $\delta \! f = f-f_{c}$.
The constants $c_{1}$ and $c_{2}$ have the form $c_{1}(d)=[\Gamma
(1-d/2)]^{2/(2-d)}$ and $c_{2}(d)=[\Gamma (2-d/2)]^{2/(2-d)}$.

\end{document}